\documentclass[acmlarge, nonacm]{acmart}

\usepackage{graphicx}

\AtBeginDocument{%
  \providecommand\BibTeX{{%
    \normalfont B\kern-0.5em{\scshape i\kern-0.25em b}\kern-0.8em\TeX}}}

\setcopyright{acmcopyright}
\copyrightyear{2018}
\acmYear{2018}
\acmDOI{10.1145/1122445.1122456}

\acmBooktitle{Woodstock '18: ACM Symposium on Neural Gaze Detection,
  June 03--05, 2018, Woodstock, NY}
\acmPrice{15.00}
\acmISBN{978-1-4503-XXXX-X/18/06}

\begin{document}

\title[Are you okay, honey?]{“Are you okay, honey?”: Recognizing Emotions among Couples Managing Diabetes in Daily Life using Multimodal Real-World Smartwatch Data}

\author{George Boateng}
\email{gboateng@ethz.ch}
\affiliation{%
  \institution{ETH Z{\"u}rich}
  \city{Zurich}
  \country{Switzerland}
}

\author{Xiangyu Zhao}
\email{xiangyu.zhao@tum.de}
\affiliation{%
  \institution{TU M{\"u}nich}
  \city{Munich}
  \country{Germany}
}

\author{Malgorzata Speichert}
\email{mspeichert@student.ethz.ch}
\affiliation{%
  \institution{ETH Z{\"u}rich}
  \city{Zurich}
  \country{Switzerland}
}

\author{Elgar Fleisch}
\email{efleisch@ethz.ch}
\affiliation{%
  \institution{ETH Z{\"u}rich}
  \city{Zurich}
  \country{Switzerland}
}
\affiliation{%
 \institution{University of St. Gallen}
 \city{St. Gallen}
 \country{Switzerland}
}

\author{Janina L{\"u}scher}
\email{janina.luescher@psychologie.uzh.ch}
\affiliation{
    \institution{University of Z{\"u}rich}
    \city{Zurich}
    \country{Switzerland}
}

\author{Theresa Pauly}
\email{theresa.pauly@psychologie.uzh.ch}
\affiliation{
    \institution{University of Z{\"u}rich}
    \city{Zurich}
    \country{Switzerland}
}

\author{Urte Scholz}
\email{urte.scholz@psychologie.uzh.ch}
\affiliation{
    \institution{University of Z{\"u}rich}
    \city{Zurich}
    \country{Switzerland}
}

\author{Guy Bodenmann}
\email{guy.bodenmann@psychologie.uzh.ch}
\affiliation{%
  \institution{University of Zurich}
  \city{Zurich}
  \country{Switzerland}
 }

\author{Tobias Kowatsch}
\email{tkowatsch@ethz.ch}
\affiliation{%
  \institution{ETH Z{\"u}rich}
  \city{Zurich}
  \country{Switzerland}
}

\affiliation{%
 \institution{University of St. Gallen}
 \city{St. Gallen}
 \country{Switzerland}
}

\renewcommand{\shortauthors}{Boateng et al.}

\begin{abstract}
Couples generally manage chronic diseases together and the management takes an emotional toll on both patients and their romantic partners. Consequently, recognizing the emotions of each partner in daily life could provide an insight into their emotional well-being in chronic disease management. Currently, the process of assessing each partner’s emotions is manual, time-intensive, and costly. Despite the existence of works on emotion recognition among couples, none of these works have used data collected from couples’ interactions in daily life.  In this work, we collected 85 hours (1,021 5-minute samples) of real-world multimodal smartwatch sensor data (speech, heart rate, accelerometer, and gyroscope) and self-reported emotion data (n=612) from 26 partners (13 couples) managing diabetes mellitus type 2 in daily life.  We extracted physiological, movement, acoustic, and linguistic features, and trained machine learning models (support vector machine and random forest) to recognize each partner’s self-reported emotions (valence and arousal). Our results from the best models — balanced accuracies of 63.8\% and 78.1\% for arousal and valence respectively — are better than chance and our prior work that also used data from German-speaking, Swiss-based couples, albeit, in the lab. This work contributes toward building automated emotion recognition systems that would eventually enable partners to monitor their emotions in daily life and enable the delivery of interventions to improve their emotional well-being.
\end{abstract}



\maketitle

\section{Introduction}
For couples in which one partner has a chronic disease such as cancer and diabetes, their relationship plays a key role in the disease management if partners share the responsibility of its management \cite{seidel2012, rintala2013}. Such joint disease management, also called dyadic coping \cite{badr2017, bodenmann1997, revenson2011} takes an emotional toll on both patients and spouses  \cite{settineri2014}. Consequently, understanding each partner’s emotion within the context of their interactions and disease management in daily life could enable the triggering of various dyadic interventions (where partners are both involved e.g. \cite{ketcher2021}) to improve their emotional well-being and chronic disease management.

However, assessing emotions among couples is challenging. Two approaches are used for emotion assessment in the lab and in daily life: self-report and observer reports. For self-reports, couples can be asked to have an emotionally charged conversation that is videotaped (e.g., in the lab), and then afterward, each partner provides emotion ratings, for example, while watching the videos \cite{roberts2007} or by using a validated affect instrument such as the PANAS \cite{watson1988}. In the case of daily life, couples are periodically asked to complete self-reports \cite{schoebi2008} such as the PANAS which can be obtrusive and impractical for continuous emotion assessment. These ratings could be biased (for example, if the partner desires to project a certain emotion rating rather than how they really feel) and may not reflect the partner’s actual emotion. For observers' reports, people are trained to watch the video recordings (e.g., in the case of lab data) and use a coding scheme to rate the emotional behavior of each partner (e.g., SPAFF \cite{coan2007}). Such coding is also done for example, for audio data collected from couples’ daily life interactions \cite{robbins2014}. This manual coding process is costly and time-consuming as multiple coders need to be trained for this task \cite{kerig2004} and suffers from inter-rater reliability issues \cite{heyman2001, metallinou2013}. Automated emotion recognition of each partner’s emotion could potentially address these limitations. Current approaches for automatic emotion recognition among couples have all used data collected from the lab \cite{boateng2022a}. There exists no system that automatically recognizes the emotions of romantic partners using real-world data from couples’ interactions in daily life. One potential reason for this gap is that collecting and processing such data is non-trivial, time-intensive, and costly \cite{boateng2022a}. 

Smartwatches have been used for mood recognition of individuals \cite{budner2017} and they could be leveraged for recognizing each partner’s emotions based on the couple’s interactions in daily life. Several features of smartwatches make them well suited for this task. They are mostly with the wearer as opposed to a smartphone which could be in various places like the pocket, or bag, and just not in proximity with the user. Also, consumer smartwatches could be used to collect a wide variety of sensor data that have been used for emotion recognition in the past: audio \cite{schuller2018}, heart rate, accelerometer, and gyroscope (for gestures e.g., \cite{schmidt2019}), and ambient light (to detect the context of couples). Multimodal fusion of these sensor data could produce better recognition results \cite{poria2017, dmello2015}. Furthermore, smartwatches could be leveraged in novel ways (using Bluetooth signal strength and voice activity detection) to specifically capture partners’ interaction or conversation moments in daily life \cite{boateng2022b} for use in emotion recognition. 

In this work, we collected 85 hours (1,021 5-minute samples) of real-world multimodal smartwatch sensor data (speech, heart rate, accelerometer, and gyroscope) and trained machine learning models to recognize each partner’s emotions. Specifically, we trained models to recognize each partner’s emotional valence (negative vs positive) and emotional arousal (high vs low) during the conversation using sensor and self-report data from German-speaking, Swiss-based couples managing type 2 diabetes in daily life. We addressed the following research questions:

\textbf{RQ1:} \textit{How well can romantic partners' emotions be recognized using multimodal real-world sensor data from daily life?}

\textbf{RQ2:} \textit{Which modality and multimodal combinations produce the best emotion recognition results?}

This work is the first to recognize the emotions of romantic partners using data collected from everyday life. Our contributions are as follows (1) collection and use of a unique dataset — real-world, multimodal smartwatch sensor data from German-speaking, Swiss-based couples (N=13 couples, n=26 participants), which is the first such dataset used in the literature for automatic recognition of partners’ emotions (2) approaches for validating and quantifying data quality on manually coded, annotated and transcribed real-world speech data  (3) development and evaluation of a machine learning system to recognize the emotions of each partner using a wide variety of sensor data — acoustic, linguistic, heart rate, accelerometer, and gyroscope  (4) an investigation of the sensor modality combinations which result in the best emotion recognition performance of romantic partners. This work is an extension of the work \cite{boateng2020d} and implements the research plan that was proposed in that work, along with machine learning experiments and reported results.

In the rest of this paper, we discuss background and related work in Section \ref{sec:related_work}, methodology in Section \ref{sec:method}, experiments and evaluation in Section \ref{sec:experiments} and results and discussion in Section \ref{sec:results}, limitations and future work in Section \ref{sec:limitations}, and we conclude in Section \ref{sec:conclusion}.

\section{Background and Related work}
\label{sec:related_work}
In this section, we describe various emotion models, multimodal emotion recognition, and works that have been done to recognize emotions among couples.

\subsection{Emotion Models}
There are mainly two models of emotions used in the literature in emotion recognition: categorical and dimensional. Categorical emotions are based on the six basic emotions proposed by Ekman: happiness, sadness, fear, anger, disgust, and surprise  \cite{ekman1997}. Dimensional approaches mainly use two dimensions: valence (pleasure) and arousal which are based on Russell’s circumplex model of emotions \cite{russell1980}. Valence refers to how negative to positive a person feels and arousal refers to how sleepy to active a person feels. Using these two dimensions, several categorical emotions can be placed and grouped into the four quadrants: high arousal and negative valence (e.g., angry), low arousal and negative valence (e.g., depressed), low arousal and positive valence (e.g., relaxed), and high arousal and positive valence (e.g., excited) \cite{russell1980}.

\begin{figure}
\centering
\includegraphics[width=0.5\linewidth]{{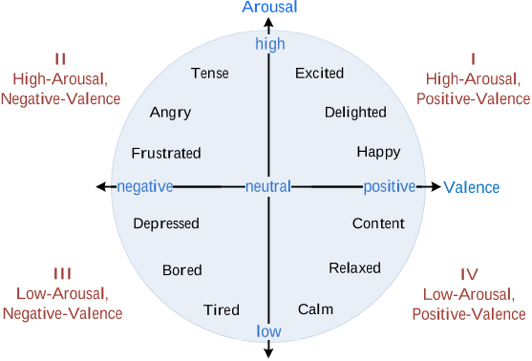}}
\caption{Russell’s circumplex model of emotions\cite{yu2016}}
\label{fig:circumplex_model}
\end{figure}

\subsection{Multimodal Emotion Recognition}
Multimodal fusion entails combining data collected from various modalities and leverages the idea that data contained in different modalities could provide a better understanding of a certain context. Various works have employed multimodal fusion approaches for emotion recognition and they have been shown to give better results than unimodal approaches \cite{poria2017, dmello2015}. There are two main fusion approaches — fusion at the feature level (early fusion) and at the decision level (late fusion). Early fusion entails combining features from different data modalities, for example, through concatenation and feeding them into the same machine learning algorithm. For late fusion, a separate algorithm is used for each data modality and then the predictions of the individual algorithms are combined using, for example, majority voting. Additional approaches include some hybrid of early and late fusion \cite{wollmer2013} and model-level fusion which leverages interactions between different modalities at the model level e.g \cite{huang2020}.

\subsection{Emotion Recognition among Couples}
Emotion recognition among couples is the task of recognizing the emotion of each romantic partner based on the context of their interaction /conversation \cite{boateng2022a}. Specifically, it entails recognizing each partner’s emotions for every utterance/speaker turn, every few seconds, or for the whole conversation. It differs from other kinds of emotion recognition tasks mainly by the kind of stimuli that induces emotions. Some stimuli are driving  \cite{zepf2020}, listening to music or watching a movie \cite{abadi2015}, and conversation between people  \cite{poria2019}. Couples’ emotion recognition is similar to emotion recognition tasks whose stimuli are conversations since it uses a conversational context. However, its uniqueness lies in the fact that the two interacting individuals are in a romantic relationship. Consequently, various insights from psychology about couples’ interaction dynamics could be leveraged to recognize each partner’s emotions. For example, romantic partners influence each other when interacting, and that insight has been used for couples’ emotion recognition (e.g.,  \cite{chakravarthula2018, boateng2021}).

There are several works that have developed machine learning systems to recognize the emotions among couples (see \cite{boateng2022a} for a detailed overview of the research field). Most of these works have been done by the Signal Analysis and Interpretation Laboratory (SAIL) team at the University of Southern California \cite{boateng2022a}. The works have mainly used emotion labels from external raters,  support vector machines as the algorithm,  the following three modalities — acoustic, lexical, and visual — with acoustic being the most used modality, and feature-level fusion of acoustic and lexical modalities \cite{boateng2022a}. There are other related work focused on recognizing behaviors among couples other than emotions  such as level of blame \cite{black2010,black2011}, conflict \cite{timmons2017}, and suicidal risk \cite{chakravarthula2020}.

Most of these works have used observer ratings (perceived emotions) rather than self-reports (one’s actual emotions) as labels. Consequently, the emotion recognition task essentially becomes recognizing external individuals’ perception of each partner’s emotion rather than each partner’s emotion per their own assessment. Though similar, the latter is more challenging. For observer ratings, coders are generally trained over several weeks, and various approaches are used to resolve ratings that are not in agreement and ensure the validity of the labels. Also, the self-reported emotion may not be reflected in that partner’s behavior in comparison to observer ratings which are purely based on behavioral observation. 

Also, several of these works have used data from English-speaking couples in the U.S. with a few using data from German-Speaking couples in Switzerland \cite{boateng2021, biggiogera2021} and Dutch-speaking couples in Belgium \cite{boateng2020a}. Additionally, several modalities such as physiological data, hand gestures, and body movement have not been explored. More importantly, none of these works have used data collected from couples’ interactions in daily life. Our work fills the current research gap by performing emotion recognition using multimodal real-world smartwatch data — speech, accelerometer, gyroscope, heart rate — and self-reported emotion data collected from German-speaking, Swiss-based couples.

\subsection{Emotion Recognition using Smartwatch Data}
There are a number of works that have performed emotion recognition using smartwatch data. AlHanai et al. \cite{alhanai2017} trained neural network models to recognize emotions using smartwatch and smartphone data collected from 10 subjects who told 31 personal stories (15 happy)  in a lab. They used an iPhone to collect audio which they transcribed. They also collected physiological and movement data with the smartwatch. They extracted 386 acoustic (functionals over low-level descriptors), linguistic (average positive and negative sentiment of words), physiological, and movement features (mean, median, variance of electrocardiogram, photoplethysmogram, accelerometer, gyroscope, bioimpedance, electric tissue impedance, galvanic skin response, and skin temperature) and selected 10 features for use using sequential forward features selection. They classified the whole narration as happy or sad (indicated by the subject) and 5-sec segments as positive, negative, or neutral (annotated by a research assistant with balanced distribution). Though they used naturalistic data (personal narratives), this work did not use data collected from an uncontrolled, real-world context such as in daily life. 

Budner et al. \cite{budner2017} trained Random Forest models to recognize moods using smartwatch data collected from 60 subjects in daily life. They classified 9 mood states (angry, sad, tired, excited, happy, quiet, elated, very happy, relaxed) and 3 levels of pleasure and activation. They extracted the following features related to body sensor data: vector magnitude counts (a measure of the total amount of movement), heart rate, and external influences: light level and GPS coordinates (variance), weather, (humidity, temperature, cloudiness, windiness, air pressure), the hour of the day, whether it was the weekend and day of the week.  Arano et al. \cite{arano2020} built upon the work by Budner et al. \cite{budner2017} and proposed the use of the smartwatch-based system to measure emotions in a real-world scenario: classroom. They were able to collect data from 30 subjects related to body sensor data:  accelerometer, light, audio, heart rate (from a smartwatch), GPS data (from a smartphone), and environmental variables (e.g., weather, longitude, latitude, altitude, room temperature, humidity, pressure, wind level, clouds level, noise level). Subjects indicated their level of Activation, Tiredness, Pleasance, Quality (of lecturer's presentation), and Understanding (of lecturer's presentation) on a scale from 0-2. They extracted statistical features and used 9 models: K-Nearest Neighbor, Decision Trees, Support Vector  Machines, Multilayer perceptron, logistic regression, Gradient Boost, XGBoost, and LSTM.

Kanjo et al \cite{kanjo2018} developed models to recognize emotions using body sensor and environmental data collected from the wild. They collected data from 40 females walking around Nottingham city center, UK for 45 mins: body movement, activity, heart rate, Electrodermal activities and body temperature and, environmental data including noise level (Env-noise), air pressure and ambient light levels, and GPS data. User emotions labels are collected using self-report input, based on a scale for valence (1-5). They used the Microsoft Band and Android phones (to collect noise, GPS, and self-report). They extracted 87 features: mean, median, max, min, range, and standard deviation and quartiles and selected 21 after feature selection. They trained ensemble models (stacking) to perform classification of the 5 levels of valence. They had a base model for each modality and a stacking model which fused the results of both models. They used the following models: Support Vector Machine, Random Forest, and K Nearest Neighbour as the base models, and Naive Bayes as the stacking model Learner which fused the base models’ predictions.

Quiroz et al. \cite{quiroz2018} developed a smartwatch-based method to recognize emotions based on movement data. They collected data from 50 subjects: (43 females; mean age 23.18 [SD 4.87] years), North-West, UK. They collected emotion data with the PANAS before and after emotion elicitation; happy, sad, and neutral. They used audiovisual movie clips and audio music clips to elicit emotions. They asked the subject to walk for 250 meters while wearing a smartwatch and heart rate monitor strap on the chest. 18 were assigned to the audiovisual condition and watched the movie before walking. Out of the 32 assigned to audio, half of them listened while walking. It took 20 mins for each subject. They extracted 107 features over 1-second sliding windows with 50\% overlap over filtered signal (accelerometer and gyroscope, and heart rate). They trained personalized models from 44 subjects to classify happy vs sad and happy vs sad vs neutral. The data was balanced. They used 10-fold stratified cross-validation with logistic regression and random forest. 

Schmidt et al, 2019 \cite{schmidt2019} trained a convolutional neural network (CNN) model to predict emotions (arousal, valence), anxiety, and stress from real-world smartwatch-based physiological and motion data. They used an Empatica E4 to collect 1,400 hours of data; accelerometer, photoplethysmogram (PPG), EDA, and skin-temperature data from 12 subjects (7 male). Subjects received EMA prompts every 2-2.5 hours or triggered manually: 1) self-assessment mannequins assessing valence and arousal 2) State-Trait Anxiety Inventory (STAI) on 6 levels and 3) Stress level scored on a four-point Likert scale. The data was skewed for all the labels. They preprocessed the data resulting in 1083 valid windows/questionnaires. The data was split between 3 levels for all the labels except stress which was binarized.  They extracted 62 features (e.g., mean, standard deviation, heart rate, heart rate variability). They used leave-one-subject-out cross-validation(LOSO) and leave-target-questionnaires-out (LTQO). As baseline models, they used different tree-based classifiers (decision-tree (DT), randomized decision trees (ET), and random forest (RF)). They used a single-task and multi-task CNN which takes the raw sensor data as the main model with late fusion. 

Park et al., 2020 \cite{park2020} developed WellBeat, a smartwatch-based system for assessing the emotional well-being of individuals. They used a Samsung smartwatch to collect PPG and heart rate data from 12 subjects (3 female) continuously throughout the day: 1121 hours of data from a 3-week study (about 445 hrs eliminated) and 1032 self-report labels related to happiness, awakeness, and relaxedness levels (1-5).  Subjects were asked to complete the self-report 3 times a day at random times during waking hours. They performed data preprocessing by removing samples where the watch was not worn, partitioned data into consecutive 5-min slices, filtered out signals without heart rate signals, extracted heart rate and RR intervals, and HRV parameters such as RMSSD, and estimated their validity. The label was matched to the day -10 mins to +10 mins around the label timestamp similar to Schmidt et al (2019). They performed classification with logistic regression and 10-fold cross-validation.

Our work builds upon some of these works by using similar preprocessing approaches and  features (for heart rate, accelerometer data, speech data, gyroscope), algorithms and evaluation approach. One modality that is missing in most of these works is the linguistic modality. We leveraged recent advances in deep learning and natural language processing to extract linguistic features from speech. Also, we systematically evaluated the performance of individual modalities and various modality combinations. The key way our work differs from these works though is our use of the context of couples’ interactions in daily life to recognize each partner’s emotion.

\section{Methodology}
\label{sec:method}
In this section, we describe how we collected and preprocessed the data and the features that we extracted.

\subsection{Data Collection}
\begin{figure}
\includegraphics[width=\linewidth]{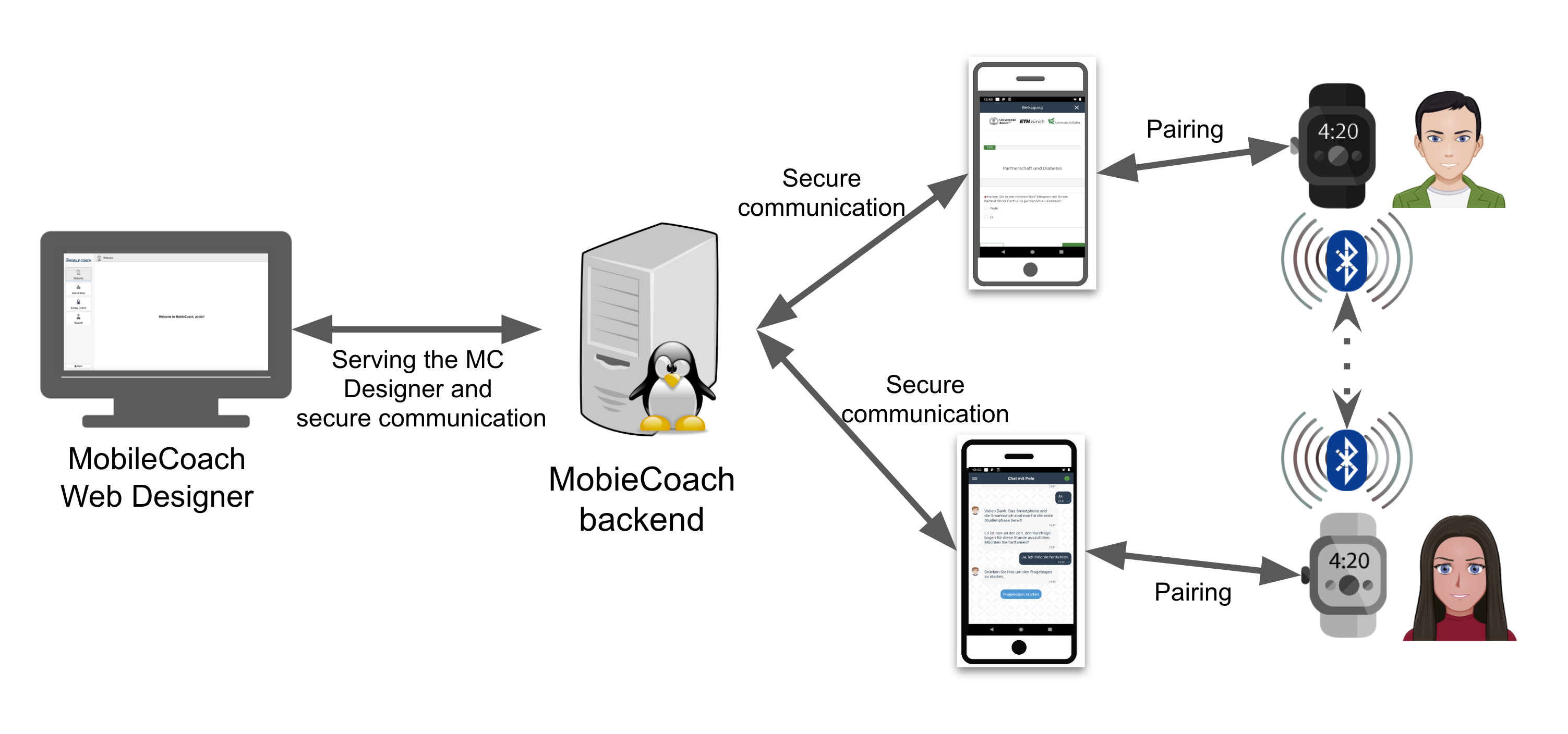}
\caption{Overview of the DyMand system} 
\label{fig:dymand_system}
\end{figure}

We developed DyMand, an open-source smartwatch and smartphone system which we used to collect data from couples in daily life in a user study (see \cite{boateng2022b} for a detailed description). The DyMand system (Figure~\ref{fig:dymand_system}) consists of a smartwatch app, and a smartphone app built on top of the MobileCoach platform \cite{filler2015, kowatsch2017} that consists of a web-based intervention designer and backend. 

We ran the DyMand study between 2019 and 2021 with heterosexual romantic couples from the German-speaking part of Switzerland in which one partner had type 2 diabetes \cite{luescher2019}. In total, we collected 85 hours of sensor and self-report data from 13 couples aged 47 to 81 years, with a mean age of 68 (SD = 9).

The study was advertised in various places including hospitals, magazines, local newspapers, and the diabetes association in Switzerland. Interested couples completed a web-based questionnaire to screen them for the inclusion and exclusion criteria, and collect socio-demographic information. Those who met the eligibility criteria were able to pick a date for a baseline assessment. During this session, both partners received comprehensive information about the study, signed the informed consent form, and completed a web-based questionnaire that captured constructs of interest at baseline that were not assessed daily. 

Each partner was given a smartwatch (Polar M600 running Wear OS) and a smartphone (Nokia 6.1 running Android 9.0), both paired and running the DyMand apps. They also received instructions on the study and then trained research assistants helped them to set up their devices and pair the corresponding smartphone and smartwatch. They were instructed to have all devices with them every day for 7 days from getting up until going to bed. To prevent mistakes from one partner accidentally using the other partner’s watch and phone, one set of phones and watches had black covers and the other set had white covers. The patient was given the white set and the supporting partner was given the black set. The partners picked the hours during which we could collect data from them. During the week, they could choose a period for the morning hours (any time between 4 am to 11 am, at least 2 hrs) and a period for the evening hours (any time between 4 pm to 11 pm, at least 2 hrs). During the weekend, data was collected all day and couples chose a start time in the early morning hours and an end time in the late evening hours (e.g., from 6 am to 10 pm). With this procedure, privacy aspects were addressed by primarily focusing on situations, in which the couples spent time together and thus reducing the number of audio recordings during the day of weekdays when chances are higher that subjects are working or moving around in public places.

We collected data from their daily life for 7 consecutive days starting the next Monday after their visit until the following Sunday night. The DyMand system triggered the collection of sensor and self-report data for 5 minutes each hour during the hours that partners pick. We collected the following sensor data from the smartwatch: audio, heart rate, accelerometer, gyroscope, Bluetooth low energy (BLE) signal strength between watches, and ambient light. 

We collected a maximum of 5 minutes of data per hour for privacy reasons. Hence, to optimize the quality of data collected within that hour and to ensure that we recorded the most relevant 5 minutes of data (when partners are interacting), rather than triggering data collection at random or scheduled times which is the norm (\cite{mehl2012, robbins2014}), the app on each of the two smartwatches collected data when 1) the partners were physically close and 2) when there was speech (see \cite{boateng2022b} for the full details). 

\begin{figure}
\includegraphics[width=0.3\linewidth]{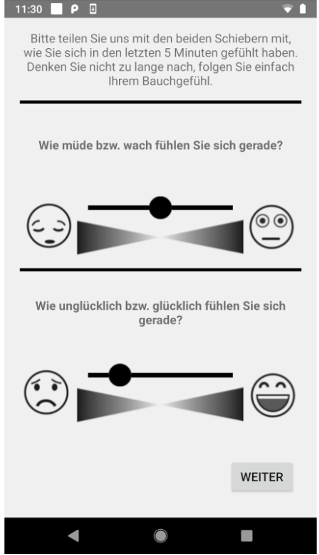}
\caption{Emotion rating with Affective Slider} 
\label{fig:affective_slider}
\end{figure}

Our algorithm used a two-step process. First, the app determines physical closeness using the BLE signal strength between the two smartwatches with one acting as the central and the other acting as the peripheral. The central smartwatch scans for the peripheral device, and checks if the signal strength between them is within a certain threshold, which corresponds to a distance estimate. If this condition is met, the app on the central device determines if the partners are speaking by using a voice activity detection (VAD) machine-learning algorithm, which is implemented on the smartwatch \cite{boateng2019a}. If these two conditions are met, the central device connects with the peripheral smartwatch, starts recording, and also sends a signal to the peripheral watch to also start recording. Consequently, each smartwatch records the same interaction, albeit, sometimes with a start delay of a few seconds on the peripheral smartwatch. It is important to note that depending on the proximity of the partners, and the presence of other individuals, parts of ongoing conversations were captured to different degrees separately on each of the smartwatches even for the same recording time period. Hence it cannot be assumed that the two recordings at the same hour and minute are exact duplicates. For example, there was a case where both partners were together with two friends with all four being in proximity, having conversations. Yet, the male partner was talking directly with the male friend, and the female partner was talking directly with the female friend, and though the two smartwatches were recorded at the same time, they captured different conversations.

In the case in which the condition of physical closeness and speaking is not met in the hour, the app triggers a backup recording in the last 15 minutes of the hour. Our evaluation showed this approach for triggering data collection to capture conversation moments between partners performed better than the backup recording \cite{boateng2022b}. The app also ensured that there were at least 20 minutes between subsequent data collection to reduce the burden of the partners completing the self-reports. 

After the 5-minute sensor data collection, the smartwatch vibrates and triggers a self-report on the smartphone for that partner to complete. The self-report asks about emotions over the last 5 minutes using the Affective Slider, a digital affect measuring tool that assesses the valence and arousal dimensions of their emotions \cite{betella2016}. In particular, they responded to “how unhappy vs. happy did you feel in the last 5 minutes?" and "how tired vs. awake did you feel in the last 5 minutes" by moving a slider from 0 to 100 on a visual scale — the Affective Slider (Figure \ref{fig:affective_slider}). If the smartwatch does not receive a message from the smartphone app within 2 mins indicating that the self-report has been started, it gives another vibration alert. If once more, within the next 2 minutes, there is still no response about the start or completion of the self-report, it implies the self-report was not completed. The self-report is then dismissed. Doing this ensured that we collected data with matching sensor and self-report samples. For privacy reasons, the app deletes that audio sample if the self-report is not completed and attempts to trigger another sensor data collection and self-report later in the hour, optimizing for the case detection partner’s interactions. Other sensor samples are still kept which could result in several sensor samples per hour without audio. If a backup recording is done, which implies that it was the last recording in that hour, the audio is not deleted even if the self-report is not completed. Doing this ensures we have at least one audio recording per hour. This approach resulted in a significant number of sensor recordings without labels. At the end of the day, the system triggered the Affective Slider, and also a short form of the PANAS self-report \cite{watson1988} for the couples to report their emotions over the whole day.

There are significant ethical and privacy concerns of such a system and study since we collect audio which is sensitive data, and more so in the context of couples’ interactions with the likelihood of speech about private topics. We took several measures as follows. First, our study received ethical clearance from the cantonal ethics committee of the Canton of Zurich, Switzerland (Req-2017\_00430). Second, we ensured that we collected a maximum of 5 minutes of audio per hour in order not to record a significant percentage of the couples’ everyday life. Consequently, even if the system triggered multiple recordings in the hour, the app always deleted all but the last one before the end of the hour. Third, to protect the privacy of subjects not taking part in the study, we asked subjects to wear a tag that we give them to indicate to others around that recording may be happening and that they may be recorded. Finally, after subjects returned their devices, we gave them the option to listen to and request the deletion of any audio samples without any explanation before the study team could listen to the audio files. Similar measures have been used in previous studies \cite{mehl2012, robbins2014} and have proven adequate to safeguard the privacy of study subjects and others not taking part in the study.

\subsection{Data Annotation, Transcription and Coding}
\begin{figure}
  \centering
  \includegraphics[width=\linewidth]{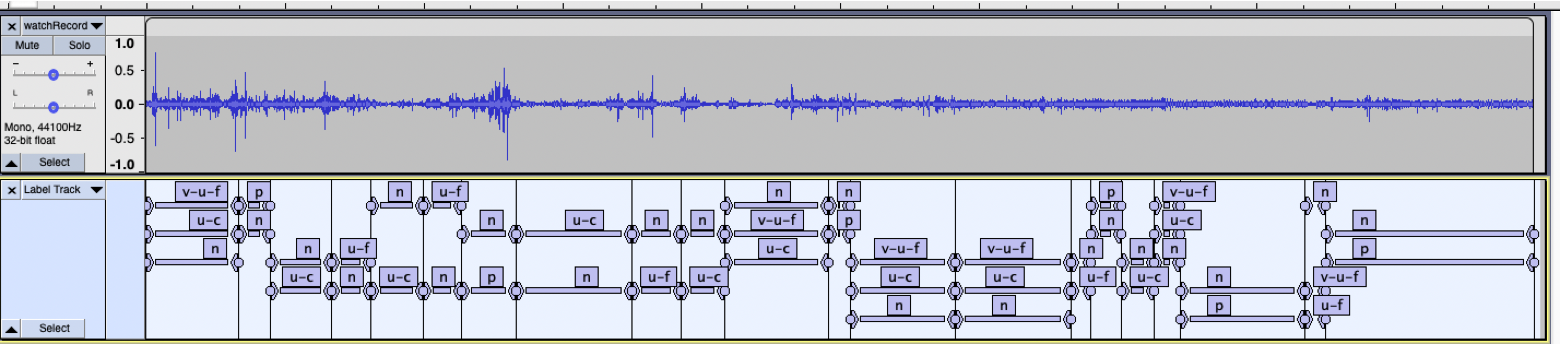}
  \caption{Screenshot of the annotation process of the audio}
  \label{fig:audacity_annotation}
\end{figure}

Four trained research assistants (RA) annotated, transcribed, and coded the audios. Using the software Audacity, each 5-minute audio was annotated with the start and end times of the speaker turns of each partner (m, f), unknown speakers (u), cross-talk between partners (c), vocalizations such as laughs, sighs (v) and the context (e.g., TV, radio), silence with no one speaking (p), noise such as music, movements of the watch, vehicles, etc. (n) and speech from radio or tv (u-tv/radio) (Figure~\ref{fig:audacity_annotation}). 

The speech of both partners within each audio was transcribed in separate documents. In particular, each Microsoft Word document was used for each partner and RAs wrote the transcript in 15 secs chunks with “//”  to separate the chunks. RAs wrote the German equivalent of any Swiss German words that were used since Swiss German is not a written language and there are different dialects of Swiss German spoken in Switzerland. Words that were not intelligible were written as “XY” in the document.

\begin{figure}
  \centering
  \includegraphics[width=\linewidth]{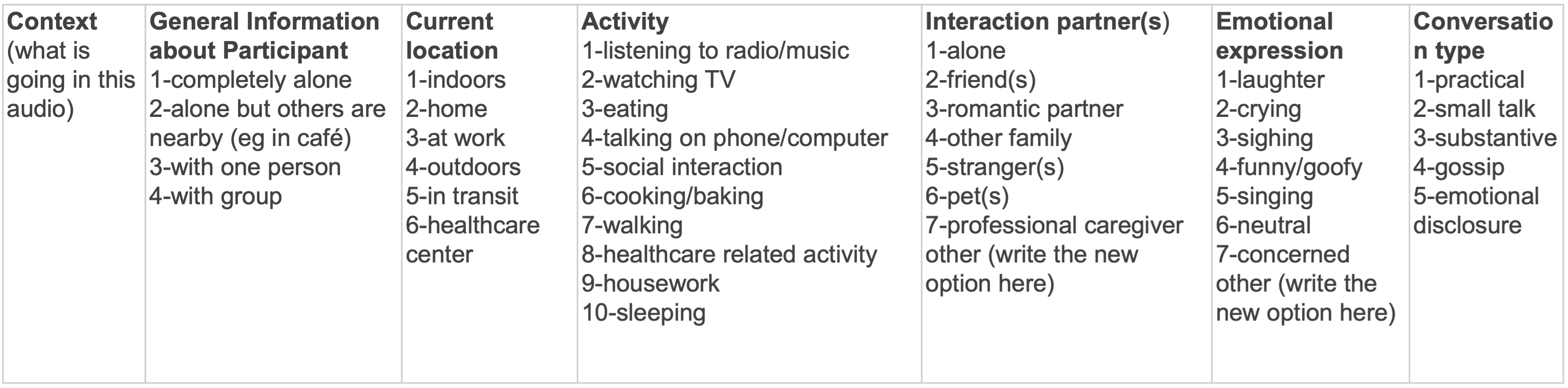}
  \caption{Screenshot of the coding context options}
  \label{fig:coding_context}
\end{figure}

RAs coded the context of each audio in a spreadsheet as they listened to the audio using a protocol based on Mehl et al. \cite{mehl2012}. They indicated if the audio contained speech,  each partner spoke, and there was a conversation and a conversation between both partners. They also provided information about the conversational context (what was going on in the audio), location, interaction partners, conversation type, activity, and emotional expression (Figure~\ref{fig:coding_context})

The real-world nature of the data posed challenges for our RAs. There were cases where the partners were having conversations with friends which made it difficult to distinguish the voices. For other cases, one partner was far away from the smartwatch while speaking, making it difficult to hear their voice. Given the challenging nature of the annotation, transcription, and coding tasks using real-world data (85 hours of audio) and the susceptibility to error, we implemented several manual and automatic approaches to perform sanity checks. We reviewed the codes to make sure the entries for different fields were consistent. For example, we cross-checked that if it is indicated that both partners spoke, then the field “interaction partner” should be “romantic partner”. Also, for each audio file, we automatically checked if there existed a non-empty transcription file and an annotation file with ‘m’ or ‘f’ if there was a “yes” for “male spoke” or “female spoke”. We also verified the accuracy of the annotations by automatically checking that for each 15-second chunk in the transcript file that contained text for the male or female partner, there existed an ‘m’ or ‘f’ in the corresponding 15-sec time period in the annotation file. We computed a percentage overlap for ‘m’ or ‘f’ with the corresponding transcript as a proxy for the quality of the annotation of that audio. Furthermore, we computed the percentage of “XY”s — inaudible words — for each audio that had speech which was a proxy for the quality of the audio and difficulty of the transcription task for that audio. RAs were given a list of files that failed these checks to then fix.

\subsection{Data Preprocessing}
We had a total of 612 self-report samples consisting of arousal and valence ratings (0 - 100) of each partner collected using the Affective Slider after the sensor data collection. We had a total of 1021 5-minute samples of sensor data (85 hours) consisting of audio, heart rate, accelerometer, gyroscope, and ambient light collected from each partner’s smartwatch. Some of the 5-minute samples were without audio data since the data collection protocol resulted in the deletion of audio samples without completed self-reports due to privacy reasons. Furthermore, because of software errors, a few of the sensor data were collected outside the data collection window and some audios were corrupted and hence could not be played or processed. We inferred these audios by eliminating audios whose size was smaller than the expected file size for 5-minute audio. Consequently, we automatically selected 5-minute samples that met the following conditions: 1) had both audio (non-corrupted) and other sensor data, and 2) were within the data collection hours specified by the partners. In total, that resulted in 1014 5-minute samples. Given our task is a supervised learning task, we also filtered and selected sensor samples that had a corresponding completed self-report.

\begin{figure}
  \centering
  \includegraphics[width=\linewidth]{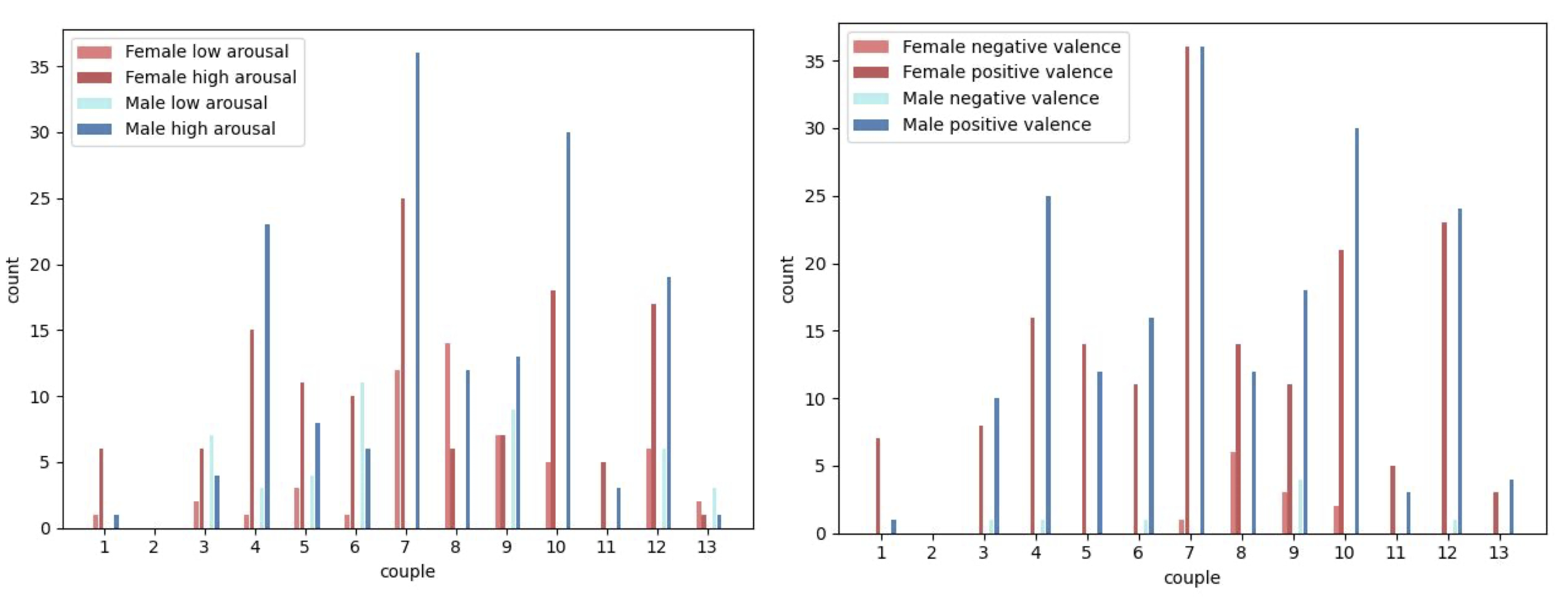}
  \caption{Distributions (bar chart) of data for arousal (left) and valence (right) per couple per gender}
  \label{fig:arous_val_distrib_per_couple}
\end{figure}

\begin{figure}
  \centering
  \includegraphics[width=\linewidth]{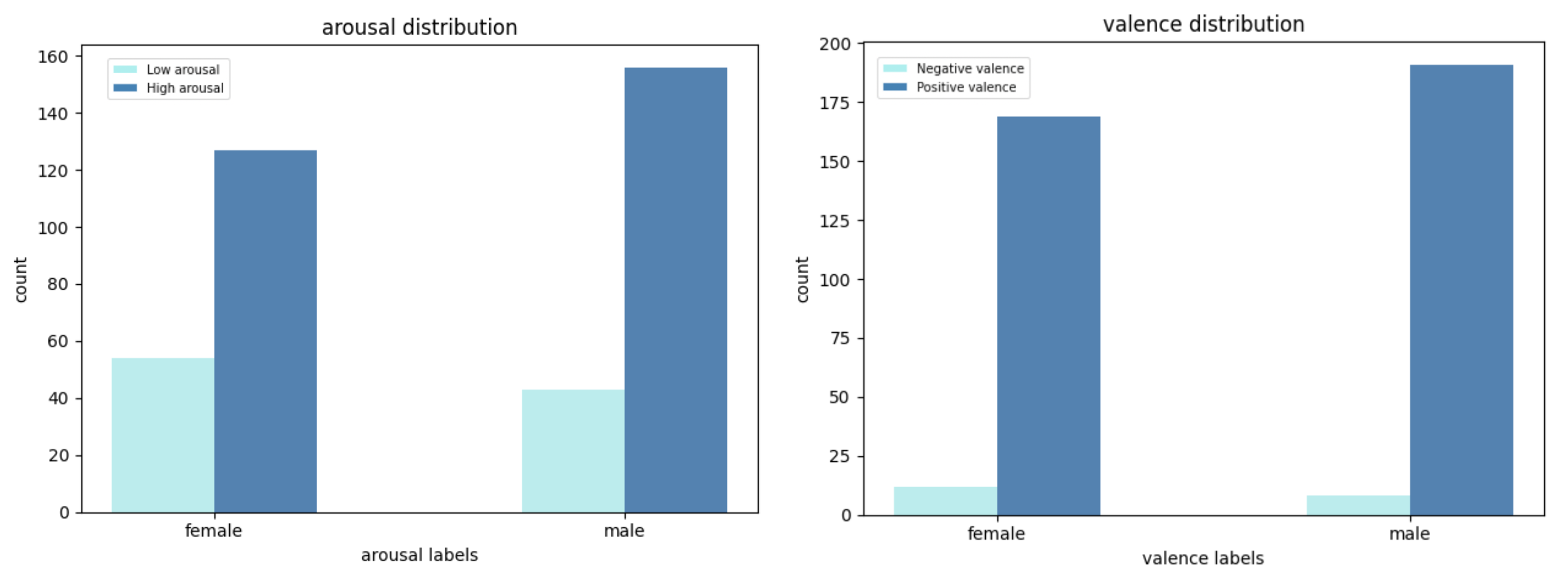}
  \caption{ Total distribution (histogram/bar chart) of data for all couples per gender for arousal (left) and valence (right)}
  \label{fig:arous_val_distrib_all_couples}
\end{figure}

We binarized the arousal and valence data into high (above 50) and low (less than or equal to 50) for arousal and negative (less than or equal to 50) and positive (above 50) for valence similar to the approach by previous works \cite{boateng2020a, boateng2021, biggiogera2021}. With binarization, the binarized arousal and valence labels can be mapped to one of the four quadrants of Russell’s circumplex model of emotions, enabling its usefulness in the real world since we can tell which group of emotions are being felt by each partner. We split at 50 because with the design of the Affective Slider, 50 was understood to be the midpoint for the labels while partners responded to it. Furthermore, taking a per subject median rating as the midpoint would be problematic if there is not a good distribution of ratings for that partner (e.g., if there are just 3 ratings which are all 80, 90, and 100, it will not be correct to assume that 80 implies negative emotions for them). Next, we filtered for samples for which we had ‘yes’ for ‘male spoke’ and ‘female spoke’ as a proxy for the context of a conversation between both partners. This filtering resulted in 380 sensor-self-report samples: 20 negative valence and 360 positive valence, 97 low arousal, and 283 high arousal. The data is highly imbalanced which is typical of real-world emotion data. Table \ref{tab:samples_per_gender} shows the sensor-self-report samples per gender. Figures \ref{fig:arous_val_distrib_per_couple}, and  \ref{fig:arous_val_distrib_all_couples} show the distributions of low and high arousal and negative and positive valence per couple per gender. We observe the skewness of the labels per couple with some couples’ data not containing any negative valence samples (1, 2, 5, 11, 13) and any low arousal samples (2, 11).

\begin{table}[]
\centering
\scriptsize
\caption{Number of negative valence and positive valence, and high arousal and low arousal  samples per gender after applying selection criteria to 5-minute samples 1) had both audio (non-corrupted) and other sensor data, and 2) were within the data collection hours specified by the partners.}
\label{tab:samples_per_gender}
\resizebox{0.5\textwidth}{!}{%
\begin{tabular}{|l|cc|cc|}
\hline
 & \multicolumn{2}{c|}{\textbf{Valence}} & \multicolumn{2}{c|}{\textbf{Arousal}} \\ \hline
 & \multicolumn{1}{c|}{\textbf{Negative}} & \textbf{Positive} & \multicolumn{1}{c|}{\textbf{Low}} & \textbf{High} \\ \hline
Male & \multicolumn{1}{c|}{8} & 191 & \multicolumn{1}{c|}{43} & 156 \\ \hline
Female & \multicolumn{1}{c|}{12} & 169 & \multicolumn{1}{c|}{54} & 127 \\ \hline
\textbf{Total} & \multicolumn{1}{c|}{\textbf{20}} & \textbf{360} & \multicolumn{1}{c|}{\textbf{97}} & \textbf{283} \\ \hline
\end{tabular}%
}
\end{table}

We filtered the audio using a low-pass filter with a cut-off frequency of 4 kHz given it was collected as raw audio at a sampling rate of 44.1KHz and human speech is less than 4 kHz.  For each 5-minute data for all the modalities, we removed outliers (data points that were more than 2 standard deviations from the mean). We resampled the data points at 50 Hz for accelerometer and gyroscope and 1 Hz for heart rate given that the Wear OS platform did not sample the signal at our specified frequency in the app. We additionally preprocessed the heart rate data to remove outlier samples such as heart rate values outside the normal range (30 to 200 beats per minute). We automatically inferred samples for which the watch was not worn. For each physiological and movement sample, we logged a value (between 0 and 3) that provides a confidence estimate of the wear state of the watch. If 50\% of the data points within the 5-minute sample had a value of 0 for the wear state, we marked the same as non-worn. 

\subsection{Features Extraction}
We describe the extraction of physiological, movement, context, linguistic and acoustic features.

\subsubsection{Physiological}
Similar to prior work \cite{kanjo2018}, we extracted the following statistical features from the heart rate data: mean, median, max, min,  25th percentile, 75th percentile, standard deviation, range, skewness, and kurtosis. The extraction resulted in a 10-dimensional feature vector.

\subsubsection{Movement}
Similar to prior work \cite{quiroz2018, schmidt2019}, we extracted the following statistical features from the accelerometer and gyroscope: mean, median, max, min,  25th percentile, 75th percentile, standard deviation, range, skewness, and kurtosis. For the accelerometer and gyroscope data, we computed the magnitude of the x, y, and z axes before using them to compute the features. We did this so that the orientation of the device does not affect the results. The extraction resulted in a 10-dimensional feature vector.

\subsubsection{Acoustic Features}
For speech, we used openSMILE to extract the 88 eGeMAPS features using the annotations corresponding to the sections of the audio where each partner spoke similar to prior work on couples’ emotion recognition \cite{chakravarthula2019, boateng2021, biggiogera2021}. These features have been shown to be a minimalist set of features adequate for emotion recognition \cite{eyben2015}. Researchers use the extraction of acoustic parameters from the speech signal as a method to understand the patterning of the vocal expression of different emotions and other affective dispositions and processes. They used a number of acoustic parameters, including parameters: time-domain (e.g., speech rate), frequency domain (e.g., fundamental frequency or formant frequencies), amplitude domain (e.g., intensity or energy), distribution domain (e.g., relative energy in different frequency bands). 

The use of machine learning led to the increase in the variety and quantity of acoustic features employed: basic (low-level ones) and derived (functionals). Therefore, finding relevant acoustic parameters is crucial in order to understand the mechanism of production and perception of emotions. Minimalistic standard parameters set for acoustic analysis of speech and other vocal sounds might lead to better generalization in real-world scenarios.  There are three criteria that guided the choice of parameters: the potential of an acoustic parameter to index physiological changes in voice production during affective processes; the frequency and success with which the parameter has been used in the past literature; its theoretical significance

The minimalistic acoustic parameter set contains 18 Low-level descriptors (LLD), which are grouped into frequency-related parameters, energy/amplitude-related parameters, and spectral (balance) parameters. They are smoothed over time with symmetric moving average filters 3 frames long (for pitch, jitter, and shimmering, only performed within voiced regions). The following functionals are applied:
\begin{itemize}
\item arithmetic mean and coefficient of variation (standard deviation normalized by arithmetic mean) to all 18 LDD
\item 20th, 50th, and 80th percentiles, range of 20th to 80th percentile, and mean and std of the slope of rising/falling signal parts are added to loudness and pitch 
\item arithmetic mean of Alpha Ratio, Hammarberg Index, spectral slopes from 0-500Hz and 500-1500Hz over all unvoiced segments 
\item rate of loudness peaks, mean length \& standard deviation of continuously voiced regions, mean length and std of unvoiced regions, number of continuous voiced regions per second  
\end{itemize}

The above functionals yield 62 parameters in the Geneva Minimalistic Standard Parameter Set. For extended Geneva Minimalistic Standard Parameter Set (eGeMAPS) the following are added so the final set contains 88 parameters: arithmetic means and coefficients of variation are applied to 7 additional LLD to all segments; arithmetic mean of the spectral flux in unvoiced regions, arithmetic mean and coefficient of variation of the spectral flux and MFCC 1-4 in voices regions + equivalent sound level. In evaluations, eGeMAPS was shown to be superior or equal to the GeMAPS \cite{eyben2015}. Hence, we extracted the eGeMAPS features resulting in an 88-dimensional feature vector.

\subsubsection{Linguistic Features}
We extracted linguistic features from the transcripts of the whole 5-minute interaction using a pre-trained model — Sentence-BERT (SBERT) \cite{Reimers2019} Similar to prior work \cite{boateng2021, biggiogera2021}. Sentence-BERT is a modification of the BERT architecture with siamese and triplet networks to compute sentence embeddings such that semantically similar sentences are close in vector space. Sentence-BERT has been shown to outperform the mean and CLS token outputs of regular BERT models for semantic similarity and sentiment classification tasks. Given that the transcripts are in German, we used the German BERT model \cite{germanbert} as SBERT’s Transformer model and the mean pooling setting. The German BERT model was pre-trained using the German Wikipedia dump, the OpenLegalData dump, and German news articles. The extraction resulted in a  768-dimensional feature vector.

\subsection{Unimodal and Multimodal Fusion}
We used the features of each modality separately as input for our machine learning experiments: physiological (heart rate), movement (accelerometer and gyroscope), acoustic and linguistic. We also used a multimodal approach with feature-level fusion \cite{boateng2021, tseng2018, chakravarthula2019}. We compared performance for individual modalities and various modality combinations as follows to answer our two research questions: physiological and movement,  acoustic and linguistic, and physiological, movement,  acoustic and linguistic.

\section{Experiments and Evaluation}
\label{sec:experiments}
We trained models for each gender to perform binary classification for arousal and valence. We trained separate models for each gender since gender differences affect how people express their emotions \cite{brody1993}. Hence, building gender-specific models \cite{boateng2020a, boateng2021} may benefit the emotion recognition task. Similar to prior work \cite{boateng2022a}, we performed couple disjoint cross-validation in which data from the same couples are never in both the train and test sets. This evaluation approach is a specific form of the subject independent evaluation but more robust as it accounts for the situation in which data from one partner (e.g., speech) may be contained in the data of the other partner \cite{boateng2022a}. We did not perform leave-one-couple-out cross-validation which is the most used evaluation approach in couples’ emotion recognition tasks \cite{boateng2022a}. Given that most couples did not have negative samples, using this evaluation approach could lead to inflated results since the model could just predict all positive results without any learning. Rather, we performed 3-fold couple disjoint stratified cross-validation. In this setup, we trained on two folds, perform prediction on the third fold as a test set, and repeated this process with each fold serving as the test fold. The stratification aspect ensures that the same ratio for classes is maintained in the train and test splits, guaranteeing that each test fold will have some negative samples. The predicted labels of each test fold are combined and the evaluation metric is computed. We used the metric balanced accuracy / unweighted average recall (UAR) due to data imbalance and confusion matrices to perform an evaluation of the predictions. We also performed hyperparameter tuning within the train split using 2-fold stratified cross-validation. We used the following machine learning models: random forest (RF), support vector machines — linear, and radial basis function (RBF) with the ‘weight’ hyperparameter set to ‘balanced’ to account for the class imbalance. We used a random baseline of 50\% for comparison.

\begin{table}[]
\centering
\caption{Evaluation results (balanced accuracy) for unimodal and multimodal models for arousal and valence for each gender}
\label{tab:results}
\resizebox{0.8\textwidth}{!}{%
\begin{tabular}{|lcccc|}
\hline
\multicolumn{1}{|l|}{\textbf{Modalities}} & \multicolumn{2}{c|}{\textbf{Arousal (\%)}} & \multicolumn{2}{c|}{\textbf{Valence (\%)}} \\ \hline
\multicolumn{1}{|l|}{\textbf{Unimodal}} & \multicolumn{1}{c|}{\textbf{Male}} & \multicolumn{1}{c|}{\textbf{Female}} & \multicolumn{1}{c|}{\textbf{Male}} & \textbf{Female} \\ \hline
\multicolumn{1}{|l|}{Physiological} & \multicolumn{1}{c|}{51.7} & \multicolumn{1}{c|}{54.2} & \multicolumn{1}{c|}{62.9} & 49 \\ \hline
\multicolumn{1}{|l|}{Movement} & \multicolumn{1}{c|}{\textbf{59.6}} & \multicolumn{1}{c|}{62.9} & \multicolumn{1}{c|}{58.2} & 50 \\ \hline
\multicolumn{1}{|l|}{Linguistic} & \multicolumn{1}{c|}{58.2} & \multicolumn{1}{c|}{\textbf{63.2}} & \multicolumn{1}{c|}{59.9} & \textbf{64.8} \\ \hline
\multicolumn{1}{|l|}{Acoustic} & \multicolumn{1}{c|}{54.8} & \multicolumn{1}{c|}{56.8} & \multicolumn{1}{c|}{\textbf{78.1}} & 61.5 \\ \hline
\multicolumn{5}{|l|}{} \\ \hline
\multicolumn{1}{|l|}{\textbf{Multimodal}} & \multicolumn{1}{l|}{} & \multicolumn{1}{l|}{} & \multicolumn{1}{l|}{} & \multicolumn{1}{l|}{} \\ \hline
\multicolumn{1}{|l|}{Physiological and Movement} & \multicolumn{1}{c|}{54} & \multicolumn{1}{c|}{\textbf{62.3}} & \multicolumn{1}{c|}{62.3} & 48 \\ \hline
\multicolumn{1}{|l|}{Linguistic and Acoustic} & \multicolumn{1}{c|}{\textbf{63.8}} & \multicolumn{1}{c|}{55.9} & \multicolumn{1}{c|}{\textbf{62.6}} & \textbf{64.9} \\ \hline
\multicolumn{1}{|l|}{Physiological, Movement, Linguistic and Acoustic} & \multicolumn{1}{c|}{59.7} & \multicolumn{1}{c|}{59.1} & \multicolumn{1}{c|}{59.6} & \textbf{64.9} \\ \hline
\end{tabular}%
}
\end{table}

\section{Results and Discussion}
\label{sec:results}
\begin{figure}
  \centering
  \includegraphics[width=\linewidth]{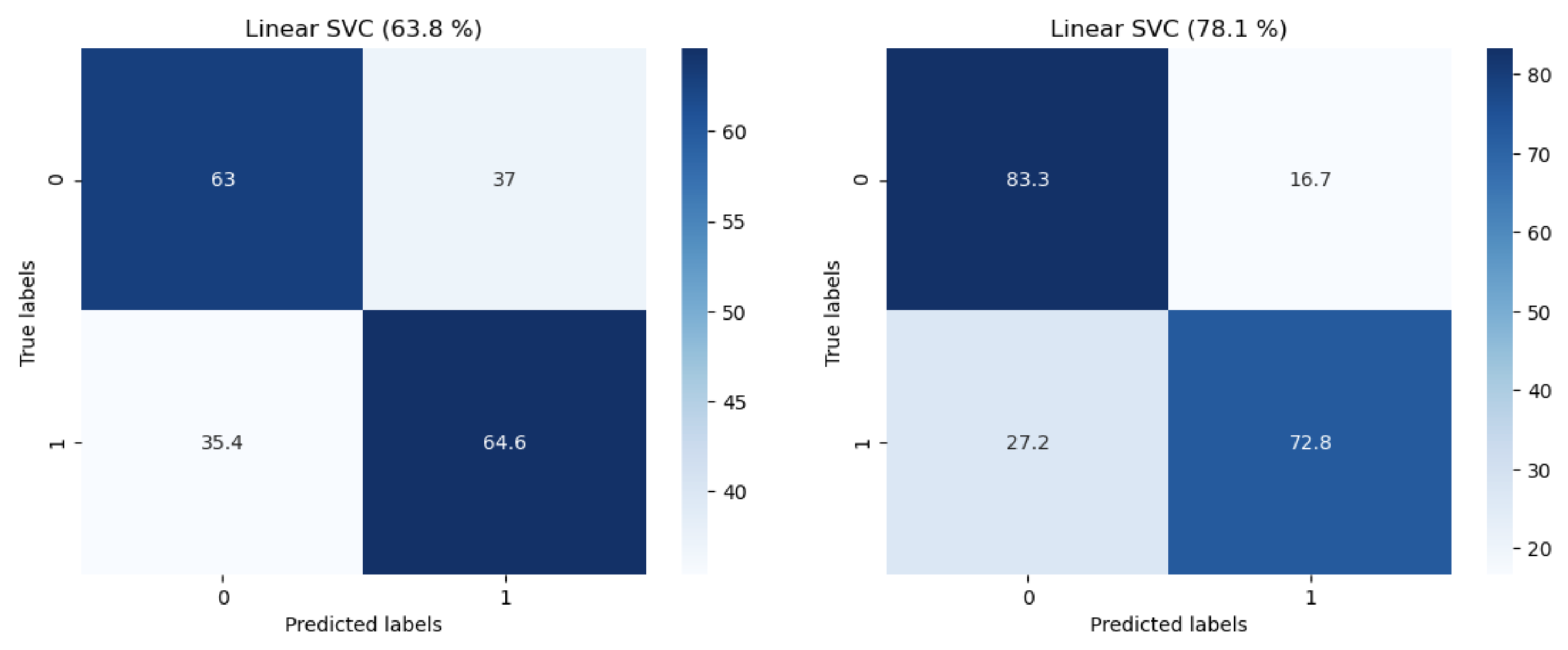}
  \caption {Confusion matrix for the best model models for arousal (left) and valence (right). Linear SVC = Linear Support Vector Machine}
  \label{fig:best_arous_val}
\end{figure}

The results of the best models are shown in Table \ref{tab:results} to answer the research questions. Among the unimodal models, for arousal, movement, and linguistic modalities performed the best for male partners (59.6\%)  and female partners (63.2\%) respectively, and for valence, acoustic and linguistic performed the best for male partners (78.1\%)  and female partners (64.8\%) respectively. Among the multimodal models, for arousal, “Linguistic and Acoustic” and “Physiological and Movement” performed the best for male partners (63.8\%) and female partners (62.3\%) respectively, and for valence, “Linguistic and Acoustic” performed the best for valence for both male partners (62.6\%)  and female partners (64.9\%), with “Physiological, Movement, Linguistic and Acoustic” also performing the same for female partners. Figure  \ref{fig:best_arous_val} shows the confusion matrices for valence’s and arousal’s best models. The linguistic and acoustic modalities produced most of the best results alone or in combination, particularly for valence, which indicates that what partners say and how they speak during their conversations are the most informative for recognizing how negative or positive they feel. This result is in line with the use of these two modalities in several couples’ emotion recognition works \cite{boateng2022a}. Furthermore, the movement modality alone or in combination with physiological modality performed the best for arousal. This result is consistent with the intuition that the greater body and hand movement are expected the more active a person feels — the arousal dimension of emotion.

We compare our results to the best results of our prior work that similarly performed global emotion recognition of positive vs negative valence from German-speaking, Swiss-based couples, albeit, with lab data. The best results (UAR) were 64.8\% (female) and 56.1\% (male) using fusion of acoustic and linguistic modality \cite{boateng2021}. Our best valence results of 64.9\% (female) and 78.1\% (male) outperform that work, albeit, only slightly for female partners. Also, as a reference, the partner-perceived emotion results reported in Boateng et al. \cite{boateng2020a} that indicate how well partner A could tell the emotions of their partner B were 73.2\% (for male partners) and 74.3\% (for female partners). We did not collect such perception data from partners in this work. Hence, a direct comparison is not possible. Nonetheless, it is worth noting that our results for male partners slightly outperform female  partners’ perceptions of their male partners’ emotions from that work.

There are significant privacy concerns with a system that recognizes the emotion of romantic couples especially considering such interactions have a high tendency to entail sensitive information. We argue that a smartwatch-based emotion recognition system such as ours has a better potential to be privacy-preserving. Compared to a facial-based emotion recognition system which can infer people’s emotions without their consent (e.g., via CCTV camera), the smartwatch system can be designed to work only with the consent of the partner. For example, the device would need to be worn to be able to collect the relevant physiological and movement data needed for emotion recognition. The speech processing component would need to work only for the speech of the partner and hence would require a speech sample from that partner to work. Furthermore, the system using partner A’s smartwatch could be designed to require a confirmation from partner B to allow their watch to share data (e.g., BLE signal or physiological data) which is needed to be able to recognize partner B’s emotion from partner A’s smartwatch. Hence, partner A could be prevented from recognizing the emotion of partner B without their consent. Furthermore, all processing of signals could be implemented to run on the device further restricting potentially sensitive data such as audio from ever leaving the smartwatch.


\section{Limitation and Future Work}
\label{sec:limitations}
The biggest limitation of this work is how highly skewed the data is especially for valence. The count for negative valence is 20 vs positive valence which is 360 with 5 out of the 13 couples having no negative valence labels. There is a self-selection bias for this study which may have resulted in having couples that are less likely to have negatively rated interactions. Future studies could target couples in therapy who may have more negative conversation moments. Furthermore, collecting data for longer than 7 days may potentially capture conversation moments with negative emotion ratings.

Additionally, data was collected from only 13 couples (26 partners). Though small, for reference, two of the most popular public emotion datasets used in emotion recognition works — IEMOCAP \cite{busso2008} and MSP Improv \cite{busso2016} — contain data from 10 individuals (12 hours) and 12 individuals (9 hours) respectively, all collected from actors in the lab. Hence, our dataset has a greater variety with reference to subjects.

Future work would explore other fusion approaches such as decision-level fusion or some hybrid approach, multitask learning since the prediction entails two target variables — valence and arousal —, and pretraining on a related dataset and then fine-tuning on this dataset. Also, further work is needed to understand the conditions under which the model performs poorly e.g., indoors vs outdoors, when the partners are together alone or with other individuals. Such analysis could provide insight into potential changes that could improve the results (e.g., additional preprocessing of the dataset). It is also critical to better understand the conditions that could degrade performance before deploying for use in the real world.

Given that we had several audio samples without labels, we had three research assistants code all the audios with emotion labels so we could have more labeled data. Unfortunately, the inter-rater agreements were poor with an intraclass correlation coefficient of 0.21 — average for all the emotions coded. The poor agreement further demonstrates the difficulty even for humans in recognizing the emotions of romantic partners. Hence, we did not use those labels for our emotion recognition experiments. This agreement could potentially become better by improving the quality of the annotation instructions and having several rounds of annotation to ensure consistency in the annotation.

Our emotion recognition system used manual speaker annotations and transcription data. Hence, there are several steps needed in the future for this system to be usable in the real world such as implementing an automatic speaker diarization (detecting when each person spoke) and a speech recognition system. In particular, current speech recognition systems do not work for this unique dataset given that the couples spoke Swiss German, which is (1) a spoken dialect and not written, and (2) varies across different parts of the German-speaking regions of Switzerland. Hence, further work is needed to develop automatic speech recognition systems for Swiss German. Also, the machine learning system needs to be implemented on the smartwatch and evaluated in real-time in the real world. The pipeline of preprocessing, feature extraction, and machine learning classification would have to be implemented using libraries and frameworks that run on smartwatch platforms such as Google’s Wear OS and or Apple’s Watch OS. Then, the system would need to be validated in a field study to evaluate the algorithm in a new, unseen context.

\section{Conclusion}
\label{sec:conclusion}
In this work, we trained machine learning models to predict the emotions of romantic partners using multimodal smartwatch data collected from daily life. We used the following sensor data: heart rate, accelerometer, gyroscope, and ambient light. We performed binary classification of valence and arousal using linear SVM, RBF SVM, and random forest. We used individual modalities and explored various combinations of modalities using feature-level fusion. Our results from the best models — balanced accuracies of 63.8\% and 78.1\% for arousal and valence respectively — are better than chance and our prior work that also used data from German-speaking, Swiss-based couples, albeit, in the lab. This work contributes toward building automated emotion recognition systems that would eventually enable couples to monitor their emotions in daily life and enable the delivery of interventions to improve their emotional well-being. This approach could also be useful in couple therapy.

\begin{acks}
We are grateful to Prabhakaran Santhanam for assisting with the study's data collection and the following research assistants for coding, annotating and transcribing our dataset: Kwabena Atobra, Elena Luzi, Denis Adamec, and Luljeta Isaki. Funding was provided by the Swiss National Science Foundation (CR12I1\_166348/1).
\end{acks}

\bibliographystyle{ACM-Reference-Format}
\bibliography{refs}

\end{document}